\documentclass[twocolumn,showpacs,preprintnumbers,amsmath,amssymb]{revtex4}
\usepackage{graphicx}
\begin{document}
\title{Gauge Fields, Geometric Phases, and Quantum Adiabatic Pumps
      }
\author{Huan-Qiang Zhou$^{1*}$, Sam Young Cho$^2$, and Ross H. McKenzie$^{1,2}$}

\affiliation{$^1$Centre for Mathematical Physics,
         The University of Queensland, 4072, Australia}
\affiliation{$^2$Department of Physics, The University of Queensland,
             4072, Australia}
\date{\today}

\begin{abstract}
 Quantum adiabatic pumping of charge and spin
 between two reservoirs (leads) has  recently
 been demonstrated  in nanoscale electronic devices.
 Pumping occurs when system parameters
 are  varied in a  cyclic manner and  sufficiently
 slowly that the quantum system always remains
 in its  ground state.  We show that  quantum pumping has a natural
 geometric representation in terms of gauge fields
 (both  Abelian and non-Abelian)  defined on the
 space of system parameters.  We make explicit the
 similarities and differences with Berry's
 geometric phase. 
 Tunneling from a scanning tunneling microscope tip through a magnetic atom
 could be used to demonstrate the non-Abelian
 character of the gauge field.
\end{abstract}
\maketitle


Normally transport of electrical charge
is dissipative (i.e., it produces heat). However, quantum adiabatic
pumping \cite{Thouless83} provides a means in nanoscale electronic
devices to use novel quantum effects
to transport single
electrons with minimal dissipation \cite{ag99}.
Furthermore, it is also possible to pump
electron spin without pumping charge \cite{sc01,a02}.
Both charge \cite{s99} and spin \cite{wpmu03} pumping have
been recently achieved experimentally, by cyclic variation of
the gate voltages that control the shape of an open
quantum dot.
This motivated extensive theoretical research in this topic,
especially on quantum charge pumping
\cite{zsa99,saa00,ak00,ae001,mm01}.
Quantum spin pumping opens
the way for applications in spintronics.
It is sometimes suggested, but not explicitly shown, that
quantum pumping is related to Berry's phase.
As first emphasized
by Berry \cite{b84}, discrete quantum systems
have the counter-intuitive
property that when some of the parameters controlling
the system are slowly varied and brought
back to their initial value the quantum state of the
system is different to the initial state.
That is,
a quantum state may acquire a geometric phase
$\exp (i \gamma _C)$ in addition to the normal dynamic phase
$\exp (-(i/ {\hbar}) \int E(t) dt)$.
This remarkable discovery may be recast
in the language of holonomy theory \cite{s83}.
Subsequent work showed
that non-Abelian gauge potentials can arise as
a result of degeneracies of energy levels of the system
\cite{wz84,msw86,z88}.
However, it is unlikely that adiabatic pumping, characteristic of
quantum {\it open} systems considered in this Letter, 
results from Berry's phase
for {\it closed discrete} systems.

We present a systematic treatment of
quantum adiabatic pumping in open systems in terms of parallel transport
and gauge fields
(both Abelian and non-Abelian) defined on the system
parameter space, 
which reveals a unifying concept of geometric phase underlying scattering
states. 
We make explicit
the similarities and differences with 
Berry's phase associated with cyclic variations of
closed quantum systems (both degenerate and non-degenerate)
(see  Table   1).
 In the scattering approach developed by Brouwer \cite{b98},
 based on an earlier work of B\"uttiker, Thomas,  and Pr\^etre \cite{btp94},
 a compact
formula was presented for the pumped charge (current) in terms of the parametric derivatives 
of the time-dependent scattering matrix subjected to the modulating potential.
We show that the pumped charge, given by Brouwer's
formula \cite{b98}, is essentially the geometric phase
associated with the $U(1)$ subgroup of the gauge group $U(M)$
($M$ is the number of channels in a certain lead),
whereas the non-Abelian sector $SU(M)$ describes 
the adiabatic pumping associated with the internal degrees
of freedom such as spin.
Expressions are given for the gauge potentials
associated with tunneling from an STM (scanning tunneling microscope)
through a magnetic atom.
We suggest an experiment 
which can be used to illustrate 
the non-Abelian character of the gauge field.

{\it The quantum system.}
Consider a mesoscopic system with $N$ leads, 
and for the $n$-th lead there are $M_n$ channels. 
Our aim is to study quantum pumping 
by periodically varying a set of the independent external parameters 
$X \equiv (X^1, \cdots, X^{\nu}, \cdots, X^p)$ 
slowly as a function of time $t$.
In the scattering approach, 
the $S$ matrix is an ${\cal N} \times {\cal N}$ matrix
with ${\cal N}$ the total number of channels, ${\cal N} =\sum _{n=1}^N M_n$.
We define vectors 
$ {\bf n}_\alpha \equiv (S_{\alpha 1}, \cdots, S_{\alpha  {\cal N}})$
in terms of the rows
of the scattering matrix $S[X(t)]$ associated with the $n$-th lead. 
The unitarity of the scattering matrix implies that these vectors 
are orthonormal 
${\bf n}^*_\alpha 
\cdot {\bf n}_\beta = \delta _{\alpha\beta}, ~~~\alpha,\beta= 1, \cdots,M_n.$
That is, this provides us with a smooth set of (local frame) bases 
${\bf n}_\alpha(t)$. 

{\it The gauge potential.}
Assume that the parallel transport law  
\begin{equation}
\Psi^*_\alpha \cdot d \Psi_\beta =0 \label {pt}
\end{equation}
holds for (fibre) vectors $\Psi_\alpha$,
where $d \Psi_\alpha$ is the variation in $\Psi_\alpha$ 
resulting from a variation  $dX$
in the external parameters.
If $\Psi_\alpha (0) = {\bf n}_\alpha(0)$, 
i.e., the initial vector describing 
the scattering process in which the incident particle comes 
from the $\alpha$-th channel in the $n$-th lead, 
then the degeneracy of the channels implies
that the transported vector $ \Psi_\alpha(t)$ 
{\it must be a linear combination of all} ${\bf n}_\alpha(t)$, 
$\Psi_\alpha(t)= \sum_\beta U_{\alpha\beta}(t) {\bf n}_\beta(t)$. 
Expressed another way, 
the transported vector describes a combined scattering
process in which particles come from all channels in the $n$-th lead.
Obviously, $U(t)$ is unitary, i.e., $U(t) \in U(M_n)$. 
Physically,
this means certain information about where the incident particles come
from is lost during parallel transport, and is
encoded in the unitary matrix $U(t)$. 
Inserting into the
parallel transport law in Eq. (\ref{pt}), we have 
\begin{equation}
(U^{-1}d U)_{\alpha\beta}= -{\bf n}^*_\beta \cdot d {\bf n}_\alpha.
\label{vpt}
\end{equation}
Since ${\bf n}_\alpha$ varies 
as the parameters $X^{\nu}$ vary with time, 
we can thus define the gauge potential 
$
 A_{\alpha\beta\nu} \equiv 
 {\bf n}^*_\beta \cdot {\partial}_\nu {\bf n}_\alpha,
 \label{eq2}
$
where $\partial_\nu \equiv \partial/\partial{X^\nu}$
so that 
$$ (U^{-1}d U)_{\alpha\beta}= -\sum_\nu A_{\alpha\beta\nu}  d X^\nu.$$
This can be integrated in terms of exponential integrals. 
For the period $\tau$ of an adiabatic cycle, we have
\begin{equation}
(U(\tau))_{\alpha\beta}= ({\rm P} \exp (-\oint 
  \sum_{\nu} A_\nu d X^\nu))_{\alpha\beta}, 
\label{u}
\end{equation}
where ${\rm P}$ denotes path ordering.
Defining $A \equiv \sum_{\nu} A_\nu dX^\nu$, 
one can see it is Lie algebra $u(M_n)$ valued and thus anti-Hermitian.
$A$ plays the role of a gauge potential, 
as in the case of Berry's phase \cite{wz84}
for closed (discrete) quantum systems. 

{\it Gauge transformation.}
The gauge group $U(M_n)$ 
originates from the unitary freedom in choosing local bases 
${\bf n}_\alpha (\alpha= 1, \cdots, M_n)$,
$$
{\bf n'}_\alpha (t) = \sum_\beta \omega _{\alpha\beta}(t){\bf n}_\beta (t).
$$
This amounts to different choices of the scattering matrix: 
$S'(t) =\Omega (t) S(t)$ with $\Omega (t)$ a diagonal block matrix, 
the $n$-th block of which is 
an $M_n \times M_n$ unitary matrix $\omega (t)$. 
Physically speaking, left multiplication of the scattering matrix $S(t)$
by $\Omega(t)$ just redistributes the scattering particles among
different incoming channels 
associated with a certain lead, which does not affect
correlations at the scatterer and so the physics remains the
same. 
The gauge potential $A(t)$ transforms as
\begin{equation}
A'(t) = d \omega \omega^{-1} + \omega A \omega^{-1}. \label{gt}
\end{equation}
The gauge field strength  defined by
$F \equiv dA - A \wedge A$   
transforms covariantly $F' = \omega F \omega^{-1}$. 
Therefore, a $U(M_n) \equiv U(1) \times SU(M_n)$ gauge field is
defined on a $p$-dimensional
parameter space which drives the quantum pumping.
The trace of $U(\tau)$ given by Eq. (\ref{u}) is gauge-invariant.

{\it Justification of parallel transport.}
Now we need to justify the assumption of the parallel transport 
law in Eq. (\ref{pt}). 
Physically, by  ``adiabatic'' we mean that 
the dwell time $\tau_p$ during which particles scatter off the scatterer
is much shorter than the time period
$\tau= 2\pi/\omega_a$ during which 
the system completes the adiabatic cycle.
Here $\omega_a$ is a slow frequency characterizing the adiabaticity and
$\tau_p$ is related to
(but not determined alone by) Wigner time delay matrix
$\tau_w(s,E) \equiv -i S^\dagger(s,E) \partial S(s,E)/\partial E $ with $E$
being the energy of scattering particles 
and $s$ being the so called {\em epoch} defined as $s=\omega_a t$ \cite{ae003}.
Then
the response to
the variation of the particle
distribution in a certain channel {\it is only limited to} 
channels associated with the same lead. 
That means we ignore any responses which involve channels
in different leads. 
Such a response can be treated as dissipation,
a correction to the adiabatic limit. Then, the parallel transport law 
in Eq. (\ref{pt})
follows from the parallel transport law for the wave function.
The latter is a solution of the time-dependent Schr\"odinger equation
$i \hbar \partial_t |\Phi(t)\rangle = H(t) |\Phi(t) \rangle$. 
As is well known, the Schr\"odinger equation induces a parallel transport law
${\rm Im}\langle\phi(t)|\partial_t \phi(t)\rangle=0$ \cite{aa87}, 
with $|\phi(t)\rangle \equiv \exp (i\int h(t)dt)
|\Phi(t)\rangle$, where 
$h(t) = \langle\Phi|H|\Phi\rangle/\langle \Phi|\Phi\rangle$. 
We can write the wave function $|\phi(t)\rangle $ as a linear combination of
{\it all} scattering states associated with a certain lead in the adiabatic
case. 
Formally, $|\phi(t)\rangle = \sum _\alpha c_\alpha |\psi_\alpha(t)\rangle$, 
with $|\psi_\alpha(t) \rangle $ denoting the scattering states
in which the scattered particles come from channels associated with the
$n$-th lead,
and $c_\alpha$ being arbitrary constants.
Then we have $\langle \psi_\alpha(t) | \partial_t \psi_\beta (t) \rangle=0$.
The adiabatic assumption implies that
$|\psi_\alpha(t)\rangle$ may be expanded in terms of instantaneous 
asymptotic scattering states, 
$|\psi_\alpha(t)\rangle 
 = \sum_\beta U_{\alpha\beta}(t) |\psi^S_\beta(t)\rangle$,
with
$|\psi^S_\alpha(t)\rangle 
= |\alpha\rangle_{in}+\sum^{\cal N}_{\beta=1} S_{\alpha\beta}(t) 
  |\beta\rangle_{out}$,
$\alpha=1, \cdots, M_n$.
Here, $|\alpha \rangle _{in}$ and $|\beta \rangle_{out}$ denote, 
respectively, the incoming and outgoing scattering states,
which are normalized such that they carry a unit flux. 
Substituting into
the parallel transport law for $|\psi_\alpha\rangle$, 
one gets Eq. (\ref{vpt}) which is equivalent to the parallel transport law
for row vectors of the scattering matrix.

{\it Quantum adiabatic pumping.}
In order to establish the connection between the geometric phase
above and the quantum pumping charge,
we need to consider the time-reversed scattering states
$|\hat\psi^S_\alpha(t)\rangle 
= |\hat{\alpha}\rangle_{in}
 +\sum^{\cal N}_{\beta=1} S_{\beta\alpha}(t) |\hat{\beta}\rangle_{out}$
with $\hat{}$ denoting the counterparts under time reversal operation
\cite{note},
which constitute a solution of the Schr\"odinger equation for
the time-reversed Hamiltonian $\hat{H}$ at any given (frozen) time
{\it at the epoch scale} \cite{ae003}.
This gives rise to another gauge potential 
$\hat{A}_{\alpha\beta\nu}
 \equiv \hat{\bf n}^*_\beta \cdot \partial_\nu \hat{\bf n}_\alpha$
with $\hat{\bf n}_\alpha \equiv (S_{1\alpha}, \cdots, S_{{\cal N} \alpha})$,
i.e., the column vectors of the scattering matrix $S(t)$.
In this case, the gauge group arises from
redistribution of scattering particles among different outgoing channels.
If $\hat {\bf n}'_\alpha(t) = \sum_\beta \hat \omega_{\alpha\beta}
   \hat {\bf n}_\beta(t)$, then the gauge transformation takes
$\hat A'(t) = d \hat \omega \hat \omega^{-1} 
+ \hat \omega \hat A \hat \omega^{-1}$.
The gauge fields $A$ and $\hat A$ are connected via time reversal operation.
If we identify the emissivity into 
the $\alpha$-th channel in the $n$-th lead as 
${\rm Im} [{\hat A}_{\alpha\alpha}/2\pi]$ \cite{btp94}, 
then we immediately reproduce
Brouwer's formula \cite{b98} describing charge pumping, 
which turns out to be associated
with the Abelian subgroup $U(1)$,
\begin{equation}
Q= \frac {e}{2\pi} {\rm Im} \oint {\rm Tr} {\hat A}, \label {pc}
\end{equation}
with $Q$ being the charge transferred 
into the $n$-th lead during one cycle. 
That is, the charge transferred 
during adiabatic pumping 
is essentially the geometric phase 
associated with the charge sector $U(1)$. 
This also explains why 
Planck's constant $\hbar$ does not  occur in the
adiabatic quantum pumped charge (current), a peculiar feature different
from the Landauer-B\"uttiker conductance. 
However, as is well known, the geometric phase
is determined only up to a multiple of $2\pi$. 
This concerns global geometric properties,
i.e., the winding number of the overall phase of 
the gauge transformation in Eq. (\ref{gt}),
$ N\equiv 1/(2\pi i) \oint {\rm Tr} (d\hat \omega \hat \omega^{-1})$.
The requirement that all physical 
observables be invariant under the gauge transformation leads 
us to the conclusion that $O(Q)=O(Q-eN)$, 
with $O$ denoting any observable. 
This result has been noticed by Makhlin and Mirlin \cite{mm01},
without proper justification, 
for the counting statistics in quantum charge pumps 
(see also, Ref. \cite{ak00}). 

To see the effects caused by  non-Abelian gauge potentials, 
we need to consider gauge invariant quantities.
Besides ${\rm Tr} \hat{U}(\tau)$, we see that both the determinant 
and eigenvalues of $\hat{U}(\tau)$ are gauge invariant. 
Actually, there are $M_n$ independent gauge invariant quantities
such as the eigenvalues $\exp[i\gamma_\alpha]$.
On the other hand, there are $M_n$ independent simultaneous
observables such as
the pumping currents $I_\alpha$ flowing into the $\alpha$-th channel, 
which must be gauge invariant.
Therefore, one may expect that
the pumping currents $I_\alpha$ are some functions of $\gamma_1, \cdots, \gamma_{M_n}$.
Because our argument only relies on gauge invariance and 
does not depend on any details of the system, such functions must be
model-independent. 
Guided by the results for the so-called
``Abelianized" non-Abelian gauge potentials, i.e.,
the gauge potentials which
turn out to be diagonal in a certain gauge, 
we have 
\begin{equation}
I_\alpha =-\frac{1}{2\pi \tau} \gamma_\alpha.
\end{equation}
This connects the physical observables with the eigenvalues of the geometric
matrix phase.
Especially, the charge pumping current $I_c$ corresponding to
the Abelian sector $U(1)$ is
$I_c \equiv \sum_{\alpha} I_\alpha =-1/(2\pi \tau) \sum_\alpha \gamma_\alpha$.
One may verify that this is consistent with Eq. (\ref{pc}) since
$\dot{Q}(t) = e I_c$ and $d \det \hat{U}(t) =-{\rm Tr}
 \hat{A}(t) \det \hat{U}(t)$. 
Alternatively, $Q= e \; {\rm Im} \ln \det U(\tau)$.
Similarly, we may define generalized ``spin" pumping currents associated with
the Cartan subalgebra of the non-Abelian sector $SU(M_n)$.
The simplest non-Abelian case $U(2)$ is relevant to 
the charge and spin pumping.

{\it Tunneling through a single magnetic spin.}
Consider the Hamiltonian 
which describes a system 
consisting of two leads coupled to a single site, the spin of which
has an exchange interaction $J$ with a magnetic spin
\cite{zb02},
\begin{eqnarray}
H &= &\sum_{k \in {L,R},\sigma} \epsilon_{k\sigma} 
        c^\dagger _{k \sigma} c_{k \sigma}
+ J \sum _{\sigma,\sigma'} d^\dagger_\sigma \Omega _{\sigma \sigma'} d_{\sigma'}
 \nonumber \\
 && \hspace*{1cm}
+\sum_{k \in {L,R},\sigma;\sigma'} 
 (V_{k\sigma,\sigma'} c^\dagger _{k\sigma} d_{\sigma'} +{\rm H.c.}).
\label {ham}
\end{eqnarray}
Here $c^\dagger_{k\sigma}$ and $c_{k\sigma}$ are, respectively, 
the creation and destruction 
operators of an electron with momentum $k$ 
and spin $\sigma$ in either the left ($L$) or
the right ($R$) lead, and $d^\dagger_\sigma$ 
and $d_\sigma$ are the counterparts of the single 
electron with spin $\sigma$ at the spin site.  
The quantity $\epsilon_{k\sigma}$ are the single 
particle energies of conduction electrons in the two leads, 
which we will assume
$\epsilon_{k\sigma} = v_F (|k|-k _F)$ 
with the convention that $v_F=1$, and the momentum is measured from
the Fermi surface for electrons in leads. 
The electrons on the spin site are connected to those in the two leads 
with the tunneling matrix 
elements $V_{k\sigma,\sigma'}$. 
For simplicity, we assume symmetric tunneling barriers between
the local spin and the leads, and only keep the spin-conserved coupling; viz.
$V_{L++}=V_{L--}=V_{R++}=V_{R--}=V$ and $V_{L+-}=V_{L-+}=V_{R+-}=V_{R-+}=0$.
The entries of the 
coupling matrix $\Omega$
take the form
$\Omega _{++} =-\Omega _{--}=\cos \theta$
and
$\Omega _{+-} = \Omega_{-+}^* =\sin \theta \exp (-i \phi)$.
The model is exactly soluble as far as the scattering
matrix is concerned. 

Once the scattering matrix is determined,
our general formalism leads us to 
the non-Abelian gauge potential,
\begin{equation}
\hat A= \hat A_\theta d \theta + \hat A_\phi d \phi,
\label{nb}
\end{equation}
where 
$\hat A_\theta \equiv \hat A^1_\theta \sigma^1 + \hat A^2_\theta \sigma^2 
  +\hat A_\theta^3 \sigma^3$
with 
$\hat A^1_\theta = i(\sin (\delta_1 -\delta _2) \cos \theta \cos \phi + 
(1-\cos (\delta_1-\delta_2)) \sin \phi)/4$,
$\hat A_\theta^2 = i(\sin (\delta_1 -\delta _2) \cos \theta \sin \phi - 
(1-\cos (\delta_1-\delta_2)) \cos \phi)/4$, and
$\hat A_\theta^3 =-i(\sin (\delta_1 -\delta _2) \sin \theta)/4$,
and
$\hat A_\phi \equiv \hat A^1_\phi \sigma^1 + \hat A^2_\phi \sigma^2 
 +\hat A_\phi^3 \sigma^3$ with
$
\hat A^1_\phi = -i(\sin (\delta_1 -\delta _2) \sin \theta \sin \phi -
(1-\cos (\delta_1-\delta_2)) \sin \theta \cos \theta \cos \phi)/4$,
$\hat A^2_\phi= i(\sin (\delta_1 -\delta _2) \sin \theta \cos \phi + 
(1-\cos (\delta_1-\delta_2)) \sin \theta \cos \theta \sin \phi)/4$,
and
$\hat A^3_\phi=-i(1-\cos (\delta_1 -\delta _2)) \sin ^2 \theta /4$.
Here $\delta_i \ (i=1,2)$ are 
the phase shifts defined by $\delta_1= -2 \tan ^{-1} (\Gamma/(k-J))$
and $\delta_2= -2 \tan ^{-1} (\Gamma/(k+J))$ with the tunneling rate
$\Gamma \equiv V^2$.

The gauge field strength $\hat F$ then takes the form
$
\hat F=-i(1-\cos (\delta_1-\delta_2)){\vec n}\cdot {\vec \sigma} d \Omega/4$.
Here 
$d \Omega = \sin \theta d \theta \wedge d \phi$ is the invariant area element 
and
${\vec n}=(\sin\theta \cos\phi, \sin\theta \sin\phi, \cos\theta)$
is the direction of the magnetic spin.
Obviously, this is just a simple rotation of the standard form 
$\hat F=-i (\alpha^2-1) \sigma^3  d\Omega /2$.
Up to a gauge transformation,
this is the same non-Abelian gauge potential,
found by Moody {\it et al.} \cite{msw86} for a diatomic molecule. 
This is
consistent with a theorem, proved in \cite{fm80},
stating that
the rotationally invariant connection on the sphere is essentially unique. 
To establish the relation between $\alpha$ and $\cos(\delta_1-\delta_2)$,
we need to calculate the gauge invariant quantity 
${\rm Tr} \hat F \wedge * \hat F$,
with $*\hat F$ being the dual of $\hat F$.
Then we have $\alpha^2=(3-\cos(\delta_1-\delta_2))/2$.
When $J=0$, i.e., in the absence of the direct exchange interaction
between electrons and the local spin, the gauge field is a pure gauge
because $\hat F=0$. Since the Pauli matrices are traceless, we have 
${\rm Tr} \hat A =0$, meaning that charge pumping is absent in the model
under consideration. That implies $\gamma_+ = -\gamma_-$.
Therefore,
the spin pumping current defined by 
$I_s=I_+-I_-$ becomes $I_s= -\gamma_+/(\pi \tau)$.  

\begin{figure}
 \vspace*{10.0cm}
 \includegraphics{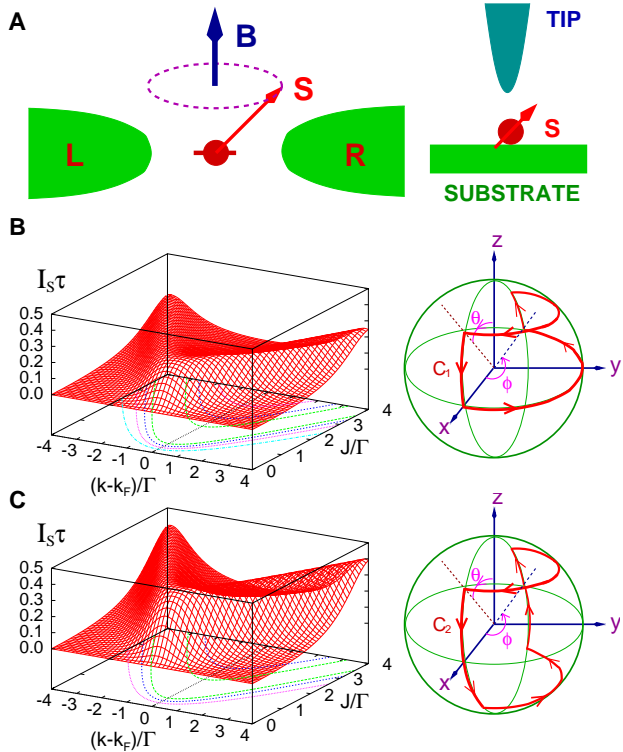}
\caption{The dependence of the spin pumping current 
        $I_s$ (times $\tau$, the period of cyclic variation)  
          on system parameters
          for two leads connected to a single magnetic spin whose direction
          is slowly varied around the path shown on the right.
          {\sf A.} Left: Schematic of the magnetic spin coupled to left (L)
             and right (R) leads.
         The magnetic spin ${\bf S}$ precesses around the direction
         of the magnetic field ${\bf B}$.
             Right: An equivalent 
             scanning tunneling microscope experimental setup.
         The pumping cycles on the parameter $(\theta,\phi)$-sphere 
         are, respectively,
         taken to be 
         $C_1 (\theta_1,\phi_1,\theta_2,\phi_2 ) = (\pi/8,0,\pi/2,\pi)$
         for {\sf B}
         and
         $C_2 (\theta_1,\phi_1,\theta_2,\phi_2 ) = (\pi/8,0,7\pi/8,\pi)$
         for {\sf C}.
         }
\label{fig1}
\end{figure}

%
 One can compute a phase factor $\hat{U}_{SR}$ which is obtained 
 from the time-reversed counterpart of Eq. (\ref{u})
 for a ``spherical rectangle (SR)''.
 From $1/2{\rm Tr} \hat{U}_{SR} = \cos \pi \tau I_s$,
 the spin pumping current $I_s$ may be extracted and 
 is shown in Fig. \ref{fig1}
 as a function of system parameters for two different paths,
 $C_1$ and $C_2$. 
 The spin pumping current takes its maximum
 value around the resonant scattering lines $k= \pm J$. 

 {\it Possible experiments.}
 A scanning tunneling microscope (STM) has been used to
 detect a quantum mirage around a single magnetic
 cobalt atom placed on a non-magnetic metallic 
 copper surface \cite{Manoharan00}.
 Electron spin resonance (ESR)-STM experiments \cite{Farle98}  
 have advanced to the point that they have spatial resolution
 at the level of a few spins \cite{Ralph01}.
 The STM setup as shown in Fig. \ref{fig1} {\sf A}
 should make it possible
 to observe gauge invariant spin pumping via a single magnetic atom
 on the surface of the substrate.
 To measure the spin pumping current $I_s$, 
 one could replace one of the leads by a ferromagnetic one.
 The spin pumping current can then be measured via the charge pumping currents.

 {\it Acknowledgments.}
  We thank Urban Lundin, Gerard Milburn, and Christoph Renner
  for valuable discussions.  
  Thanks also go to Markus B\"uttiker, Claudio Chamon,
  and Frank Wilczek for enlightening comments and encouragement.  
  This work was supported by the Australian Research Council.

 

\begin{widetext}
\end{widetext}
\begin{table}[hb]
\begin{center}
\begin{tabular}{c|c}
 \hline \hline
 Berry's Phase & Scattering (Pumping) Geometric Phase \\ \hline\hline
 Closed systems & Open systems \\
 Wave functions $|\psi_\alpha\rangle$ 
  & Row (column) vectors ${\bf n}_\alpha$ ($\hat {\bf n}_\alpha$)
     of the $S$ matrix \\
 Energy levels $E_n$ & Leads $n$ \\
 $M$ degeneracies &  $M$ channels \\
 Discrete spectrum (bound states) & Continuous spectrum (scattering states) \\
 Eigenstates in the $n$-th level & Channels in the $n$-th lead\\ 
 Parallel transport due to adiabatic theorem 
 &
 \begin{minipage}{8.5cm}
 Parallel transport due to 
 adiabatic charge and spin pumps
 \end{minipage} \\
 Gauge potential 
  $A_{\alpha\beta\nu}=\langle \psi^*_\beta| \partial_\nu 
               |\psi_\alpha\rangle$ 
  & Gauge potentials 
   $A_{\alpha\beta\nu}={\bf n}^*_\beta \cdot 
      \partial_\nu{\bf n}_\alpha$ and 
   $\hat A_{\alpha\beta\nu}=\hat {\bf n}^*_\beta \cdot 
      \partial_\nu\hat {\bf n}_\alpha$ \\
 $U(M)$ bundle & $U(M)$ bundle \\
 \begin{minipage}{5.5cm}
 Gauge group $U(M)$ arising from different choices of bases
 \end{minipage}
  & 
 \begin{minipage}{8.5cm}
 Gauge group $U(M)$ arising from redistribution of the scattering particles
 among different channels 
 \end{minipage}
 \\
 External parameters $X=(X^1, \cdots, X^p)$ 
  & External parameters $X=(X^1, \cdots, X^p)$  \\
  \hline \hline 
\end{tabular}
\end{center}
\caption{Comparison of Berry's phase and the quantum scattering (pumping)
geometric phase.}
\label{table}
\end{table}


\begin{references}
\bibitem{note}
 We emphasize that the time reversal operation here is
 carried out for the Hamiltonian at the frozen time, 
 which may be or may not be broken, depending on
 the Hamiltonian concerned.
 However, there is another time reversal operation
 at large time scale, which is always broken during pumping.
\end{references}
\end{document}